\documentclass[12pt,a4paper]{article}
\usepackage[textheight=22cm,textwidth=16.5cm]{geometry}
\usepackage{amsmath, nccmath}
\usepackage{cite}
\usepackage{amssymb}

\usepackage{dcolumn} 
\newcolumntype{d}{D{.}{.}{2}}
\newcolumntype{e}{D{.}{.}{3}}
\newcolumntype{f}{D{.}{.}{4}}
\newcommand{\fett}[1]{\mbox{\boldmath$#1$}}

\newcommand{\bra}[1]{\ensuremath{\left\langle #1\right|}}
\newcommand{\ket}[1]{\ensuremath{\left|#1\right\rangle}}
\newcommand{\braket}[2]{\ensuremath{\left\langle #1\vphantom{#2}\right.\left|\vphantom{#1}#2\right\rangle}}

\DeclareMathAlphabet{\mathpzc}{OT1}{pzc}{m}{it}
\newcommand{\mr}[1]{\mathrm{#1}}

\usepackage{color}
\usepackage{amsmath}
\usepackage{xcolor}
\newcommand{\highlight}[1]{\colorbox{gray!50}{$\displaystyle#1$}}

\makeatletter
\renewcommand*\env@matrix[1][*\c@MaxMatrixCols c]{%
  \hskip -\arraycolsep
  \let\@ifnextchar\new@ifnextchar
  \array{#1}}
\makeatother

\begin{document}

\begin{center}
~\vspace*{-0.01cm}
{\Large %
Relativistic Kinetic-Balance Condition for Explicitly Correlated Basis Functions%
}\\
\vspace{0.3cm}
{\large
Benjamin Simmen$^a$, Edit M\'atyus$^b$\footnote{corresponding author; e-mail: matyus@chem.elte.hu; present address: Department of Chemistry, University of Cambridge, Lensfield Road, Cambridge CB2 1EW, United Kingdom.}, and Markus Reiher$^a$\footnote{corresponding author; e-mail: markus.reiher@phys.chem.ethz.ch}
}\\[2ex]

$^a$ETH Z\"urich, Laboratorium f\"ur Physikalische Chemie, \\
Vladimir-Prelog-Weg 2, 8093 Z\"urich, Switzerland\\
$^b$E\"otv\"os Lor{\'a}nd University, Institute of Chemistry, P.O. Box 32, H-1518, Budapest 112, Hungary

30/07/2015\\[-0.5cm]

\end{center}

\abstract
This paper presents the derivation of a kinetic-balance condition for explicitly correlated basis functions employed in semi-classical relativistic calculations. Such a condition is important to ensure variational stability in algorithms based on the first-quantized Dirac theory of 1/2-fermions. We demonstrate that the kinetic-balance condition can be obtained from the row reduction process commonly applied to solve systems of linear equations. The resulting form of kinetic balance establishes a relation
for the $4^N$ components of the spinor of an $N$-fermion system to the non-relativistic limit, which is in accordance with recent developments in the field of exact decoupling in relativistic orbital-based many-electron theory. 

\newpage
\normalsize
\setlength{\parindent}{0cm}
\setlength{\parskip}{2ex}
\section{Introduction}
Most of relativistic quantum chemistry and molecular physics 
is based on the (first-quantized) Dirac Hamiltonian\cite{Reiher2009a,Dyall2007,Simmen2014,bary10,Hirao2004,Schwerdtfeger2002,Hess2003}. 
However, unlike its non-relativistic counterpart, the Dirac Hamiltonian is not bounded from below and measures have to be taken in order to obtain correct lower bounds for the ground- and 
excited-state energies of bound states. Depending on whether the small components of the one-fermion basis spinors are included or eliminated (by some decoupling approach \cite{Peng2012a}), 
methods are classified as four- or two-component methods.  
Four-component methods rely on the kinetic-balance condition for variational stability. This condition is well-defined for single fermions\cite{Lee1982,Stanton1984,kutz84,Ishikawa1983,Dyall1999,Pestka2004a,Kutzelnigg2007,Sun2011} and can therefore straightforwardly be applied to orbital-based methods such as the Dirac--Hartree--Fock approach and electron-correlation methods based on it\cite{Dyall1994,Jensen1996,Visscher1996,Laerdahl1997,Thyssen2001,Visscher2001,Fleig2003,Pernpointner2003,Yanai2006,Fleig2006,Knecht2010,Thyssen2008,Abe2008,knec2014}. 
For orbital-based theories with explicit correlation factors, recent work focused on four-component second-order M\o{}ller--Plesset perturbation theory with positive-energy-states projection operators in combination with the one-electron kinetic-balance condition\cite{Ten-no2012}. 
Li and co-workers have studied coalescence conditions for explicitly correlated four-component wave functions\cite{Li2012} but without addressing the issue of kinetic balance.

A first solution to the full problem of kinetic balance for explicitly correlated trial wave functions was presented by Pestka and co-workers who have published a series of papers investigating the relativistic helium-like two-electron systems 
treated as a two-electron system in a central potential\cite{Pestka1998,Pestka2004,Pestka2006,Pestka2007,Bylicki2008,pest12}. Their solution is an infinite series of transformations of the individual components of the two-electron 16-spinor which is truncated in order to obtain an approximately kinetically balanced trial wave function. Unfortunately, little technical information 
is provided in Refs.\ \cite{Pestka1998,Pestka2004,Pestka2006,Pestka2007,Bylicki2008,pest12} and it remains unclear how such an approximate kinetic-balance condition can be extended to systems containing more than two fermions. 

In this work, we extend the pioneering work by Pestka {\it et al.} on He-like atoms \cite{Pestka2007} and present a scheme which allows us to derive an explicitly correlated kinetic-balance condition 
based on row reduction and a form similar to the row-reduced echelon form of the augmented matrix. 
We begin in section \ref{sec:theory} with the presentation of the theoretical background. In section \ref{sec:kb}, we apply our scheme to a two-electron system. Then, in section \ref{sec:nrlimit}, we show that the correct non-relativistic limit is obtained. In section \ref{sec:Ncase}, 
we illustrate how the computational cost can be reduced for the $N$-fermion case by introducing systematic approximations to a given
order in the speed of light. Finally, in section \ref{sec:num}, we demonstrate the variational stability of explicitly correlated, kinetically balanced trial wave functions for 
the ground state of the He atom.

\section{Theoretical Background}
\label{sec:theory}
The relativistic description of electrons based on the many-1/2-fermion Dirac Hamiltonian provides us with a first-quantized, i.e., semi-classical formalism capturing 
essential aspects of special relativity for molecular matter\cite{Reiher2009a,Simmen2014}.

\subsection{The Relativistic Electron}
A single 1/2-fermion, such as an electron, may be described by the Dirac Hamiltonian \cite{Dirac1928,Dirac1928a}
\begin{gather}
\fett{h}_D=c\fett{\alpha}\cdot\fett{p}+\fett{\beta}mc^2+\fett{V}\ .\label{eq:onedirac}
\end{gather}
The matrices $\fett{\alpha}=(\fett{\alpha}_x,\fett{\alpha}_y,\fett{\alpha}_z)$ and $\fett{\beta}$ are defined by anti-commutation relations. 
The most common choice that respects these relations is the standard representation of 4 $\times$ 4 matrices,
\begin{gather}
 \fett{\alpha}_i=\begin{bmatrix}\fett{0}_2&\fett{\sigma}_i\\\fett{\sigma}_i&\fett{0}_2\end{bmatrix}\quad\text{with}\quad i\in\left\{x,y,z\right\}\quad\text{and}\quad
 \fett{\beta}=\begin{bmatrix}\fett{1}_2&\fett{0}_2\\\fett{0}_2&-\fett{1}_2\end{bmatrix},
\label{eq:diracMat}
\end{gather}
where $\fett{\sigma}_i$ denotes one of the three Pauli spin matrices and $\fett{1}_2$ is the two-dimensional unit matrix. $\fett{p}=(p_x,p_y,p_z)^\mathrm{T}$ is the momentum operator, $\fett{V}$ is an operator for the interaction energy due to external potentials, $m$ is the rest mass of the fermion, and $c$ is the speed of light.

It is convenient to introduce a block structure for the one-fermion eigenfunction $\fett{\psi}(\fett{r})$, the 4-spinor, 
according to the $2\times 2$ super-structure of the four-dimensional $\fett{\alpha}_i$ and $\fett{\beta}$ matrices in standard representation,
\begin{gather}
 \fett{\psi}(\fett{r})=\begin{bmatrix}\fett{\psi}^l(\fett{r})\\\fett{\psi}^s(\fett{r})\end{bmatrix}\
 =\begin{bmatrix}
| l \rangle\\
| s \rangle
\end{bmatrix}\
,\label{eq:onewf}
\end{gather}
where '$l$' denotes the so-called large and '$s$' the corresponding small component.
We refer the reader to the review by Esteban, Lewin, and S\'{e}r\'{e}\cite{Esteban2008} and the book by Thaller\cite{Thaller1992} for a more detailed mathematical discussion of the Dirac Hamiltonian and its eigenfunctions.

The spectrum of the Dirac Hamiltonian features three distinct parts. The first part comprises the discrete bound states with energies between $+mc^2$ and $-mc^2$. The second part is the positive continuum ranging from $+mc^2$ to $+\infty$. The last part of the spectrum is the negative continuum ranging from $-mc^2$ to $-\infty$.
The negative continuum is a source of instabilities in variational calculations where the Rayleigh quotient,
\begin{gather}
 E[\fett{h}_D,\fett{\psi}(\fett{r})]=\frac{\bra{\fett{\psi}(\fett{r})}\fett{h}_D\ket{\fett{\psi}(\fett{r})}}{\braket{\fett{\psi}(\fett{r})}{\fett{\psi}(\fett{r})}},
\label{eq:E}
\end{gather}
is minimized and (usually unwanted) negative-energy continuum solutions can be encountered if no precautions (such as projection onto positive-energy states) are
taken into account. For basis-set expansion techniques, Schwarz and co-workers showed that the finite size of ordinary basis sets may pose 
difficulties\cite{Mark1982,Schwarz1982}, which is therefore sometimes called the 'finite-basis disease'\cite{Schwarz1982a}.

An effective means of dealing with the problem of variational collapse is the kinetic-balance condition
\cite{Lee1982,Stanton1984,Dyall1999,Ishikawa1983,Pestka2004a,Sun2011,Kutzelnigg2007} which relates the large and the small component of the 4-spinor:
\begin{gather}
 \fett{\psi}^s(\fett{r})\approx\frac{\fett{\sigma}\cdot\fett{p}}{2mc}\fett{\psi}^l(\fett{r})\ .\label{eq:oneekb}
\end{gather}

The derivation of this relation is straightforward. The Dirac eigenvalue problem
\begin{gather}
 (\fett{h}_D-E\fett{1}_4)\fett{\psi}(\fett{r})=0
\end{gather}
leads to a set of two linear equations for the two components of the 4-spinor in Eq.\ (\ref{eq:onewf}). After the energy spectrum has been shifted by $-mc^2$ to match the
non-relativistic energy scale, this system of equations reads
\begin{gather}
 (V-E)\fett{\psi}^l(\fett{r})+c\fett{\sigma}\cdot\fett{p}\ \fett{\psi}^s(\fett{r})=0\label{eq:system1}\ ,\\
 (V-E-2mc^2)\fett{\psi}^s(\fett{r})+c\fett{\sigma}\cdot\fett{p}\ \fett{\psi}^l(\fett{r})=0\ ,\label{eq:system2}
\end{gather}
where the four-dimensional operator \fett{V} was assumed to be a diagonal matrix with the same element $V$ as diagonal entries.
We only need one of the two equations to relate the small component to the large one. Since $\fett{\sigma}\cdot\fett{p}$ has no multiplicative inverse, it is more convenient to choose the second equation in order to obtain an expression for $\fett{\psi}^s(\fett{r})$. After rearranging the terms, we obtain the exact relation for Eq.\ (\ref{eq:system2})
\begin{gather}
 \fett{\psi}^s(\fett{r})=\frac{c\fett{\sigma}\cdot\fett{p}}{(E-V+2mc^2)}\fett{\psi}^l(\fett{r})\ .\label{eq:exact}
\end{gather}
This relation depends on the energy of the system which is not known a priori but is one of the desired results of the problem. Eq.\ (\ref{eq:exact}) can therefore not be applied to our problem. 
Now, $(E-V)$ is considered small compared to $mc^2$ so that we may introduce the approximation
\begin{gather}
\label{eq:oneferm}
 E-V+2mc^2\approx 2mc^2
\end{gather}
to eliminate $(E-V)$ and arrive at the kinetic-balance condition in Eq.\ (\ref{eq:oneekb}). We note that this approximation step turned out
to be unimportant for the construction of variationally stable basis-set expansion techniques applied in four-component orbital-based theories\cite{dyal94c,Peng2012a}, which
we assume to remain valid for the $N$-particle theory to be developed in this work.

Basis-set expansions which obey Eq.\ (\ref{eq:oneekb}) provide a variationally stable parametrization of a trial wave function for a single fermion. 
Eq.\ (\ref{eq:oneekb}) may therefore be formulated in terms of the one-fermion model spaces\cite{Pestka2003,Pestka2004a,Pestka2007}
\begin{gather}
 \ket{l}\in\mathcal{H}^l\quad\text{and}\quad\ket{s}\in (\fett{\sigma}\cdot\fett{p})\mathcal{H}^l\subset\mathcal{H}^s \ .
\end{gather}

This one-fermion kinetic-balance condition can be imposed by a transformation\cite{Peng2012a}, 
\begin{gather}
 \fett{U}^{(1)}_{\mr{KB}}=
\begin{bmatrix}
U^{(1)}_{l}&0_2\\0_2&U^{(1)}_{s}
\end{bmatrix}
=
\begin{bmatrix}
\fett{1}_2&\fett{0}_2\\
\fett{0}_2&\displaystyle\frac{\fett{\sigma}\cdot\fett{p}}{p}
\end{bmatrix}
\label{eq:kbtrans}
\end{gather}
(with $p=|\fett{p}|$) 
on basis functions into which the large component of the one-fermion 4-spinor is expanded. 
Hence, the model spaces for the large and the small components are generated in terms of this transformation. 
The advantage of this form of the kinetic-balance condition is that the large-component and small-component model spaces remain normalized. 
It is also possible to transform the Dirac Hamiltonian and then form identical model spaces for the large and small components. The transformed Dirac Hamiltonian is the so-called 
modified Dirac Hamiltonian \cite{dyal94c} and is the basis of orbital-based exact-decoupling methods\cite{Peng2012a}.

The kinetic-balance condition in Eq.\ (\ref{eq:oneekb}) also ensures the correct non-relativistic (NR) limit for $c\rightarrow \infty$. 
The Rayleigh quotient of Eq.\ (\ref{eq:E}) yields in the limit $c\rightarrow\infty$ the non-relativistic Schr\"odinger energy:
\begin{gather}
 E_{\mr{NR}}=\lim_{c\rightarrow\infty}\displaystyle\frac{\bra{\fett{\psi(\fett{r})}}\fett{h}_D-\fett{\beta} m c^2\ket{\fett{\psi(\fett{r})}}}{\braket{\fett{\psi(\fett{r})}}{\fett{\psi(\fett{r})}}}
=\displaystyle\frac{\left\langle \tilde l\left|\displaystyle\frac{p^2}{2m}+V\right| \tilde l\right\rangle}{\braket{\tilde l}{\tilde l}}\ ,
\end{gather}
where $| \tilde l\rangle$ denotes the (scalar part of the) large component of the spinor after taking the limit.

\subsection{Many-Fermion Dirac Hamiltonian}

The relativistic first-quantized many-fermion Hamiltonian (with positive-energy projection not explicitly shown for the sake of brevity) reads,
\begin{gather}
\fett{H}^{(N)}_D=\sum_{i=1}^N\fett{h}_D(i)+\fett{W}^{(N)}\label{eq:nphamiltonian}
\end{gather}
with
\begin{gather}
 \fett{h}_D(i)=\fett{1}_4(1)\otimes\cdots\otimes\fett{1}_4(i-1)\otimes\fett{h}_D\otimes\fett{1}_4(i+1)\otimes\cdots\otimes\fett{1}_4(N)\ ,\label{eq:freenphamiltonian}
\end{gather}
where $\fett{h}_D$ is the one-fermion Dirac Hamiltonian of Eq.\ (\ref{eq:onedirac}) taken for fermion $i$ and $\fett{W}^{(N)}$ describes the interaction of all pairs of the $N$ fermions.
The wave function for $N$ \emph{non}-interacting fermions, i.e., $\fett{W}^{(N)}=0$, can be constructed as the direct product of one-fermion 4-spinors $\fett{\psi}_i(\fett{r}_i)$,
\begin{gather}
 \fett{\Psi}(\fett{r})=\fett{\psi}_1(\fett{r}_1)\otimes\ldots\otimes\fett{\psi}_i(\fett{r}_i)\otimes\ldots\otimes\fett{\psi}_N(\fett{r}_N)\label{eq:nonint}\ ,
\end{gather}
which can be antisymmetrized to fulfill the Pauli principle.
Now, $\fett{r}=(\fett{r}_1,\ldots,\fett{r}_N)^{\mr{T}}$ collects all $N$ one-fermion coordinates. In the case of two fermions, we have the direct product of two basis states
\begin{gather}
\fett{\psi}_1(\fett{r}_1)\otimes\fett{\psi}_2(\fett{r}_2)=
\begin{bmatrix}
\psi^{l1}_{1}(\fett{r}_1)\\\psi^{l2}_{1}(\fett{r}_1)\\\psi^{s1}_{1}(\fett{r}_1)\\\psi^{s2}_{1}(\fett{r}_1)
\end{bmatrix}\otimes
\begin{bmatrix}
\psi^{l1}_{2}(\fett{r}_2)\\\psi^{l2}_{2}(\fett{r}_2)\\\psi^{s1}_{2}(\fett{r}_2)\\\psi^{s2}_{2}(\fett{r}_2)
\end{bmatrix}
=
\begin{bmatrix}
\highlight{\psi^{l1}_{1}(\fett{r}_1)\psi^{l1}_{2}(\fett{r}_2)}\\
\highlight{\psi^{l1}_{1}(\fett{r}_1)\psi^{l2}_{2}(\fett{r}_2)}\\
\psi^{l1}_{1}(\fett{r}_1)\psi^{s1}_{2}(\fett{r}_2)\\
\psi^{l1}_{1}(\fett{r}_1)\psi^{s2}_{2}(\fett{r}_2)\\
\highlight{\psi^{l2}_{1}(\fett{r}_1)\psi^{l1}_{2}(\fett{r}_2)}\\
\highlight{\psi^{l2}_{1}(\fett{r}_1)\psi^{l2}_{2}(\fett{r}_2)}\\
\vdots\\
\psi^{s2}_{1}(\fett{r}_1)\psi^{s2}_{2}(\fett{r}_2)
\end{bmatrix}
\ .\label{eq:twokr}
\end{gather}
The superscripts '$l$' and '$s$' indicate large and small 2-spinors, respectively, as before. 
The number attached to these letters indicates the element of a 2-spinor. For instance, the elements of the large-component 2-spinor are denoted as
\begin{gather}
 \psi^{l}_{1}(\fett{r}_1)=\begin{bmatrix}
\psi^{l1}_{1}(\fett{r}_1)\\\psi^{l2}_{1}(\fett{r}_1)
\end{bmatrix}\ .
\end{gather}

A basis-set expansion of an $N$-fermion wave function may be constructed to be consistent with the model space
\begin{gather}
 \mathcal{H}^{(N)}=\mathcal{H}^{l\ldots l}\oplus\ldots\oplus\mathcal{H}^{\lambda_1\ldots \lambda_N}\oplus\ldots\oplus\mathcal{H}^{s\ldots s},
\label{eq:modelspace}
\end{gather}
where each $\mathcal{H}^{\lambda_1\ldots \lambda_N}$ is constructed from the one-fermion model spaces,
\begin{gather}
 \mathcal{H}^{\lambda_1\ldots \lambda_N}=\mathcal{H}^{\lambda_1}\otimes\ldots\otimes\mathcal{H}^{\lambda_N},
\label{eq:lambda}
\end{gather}
with $\lambda_1,\ldots,\lambda_N\in\left\{l,s\right\}$. The highlighted spinor components in Eq.\ (\ref{eq:twokr}) are those contained within the model space $\mathcal{H}^{ll}$. We recognize that the wave function in Eq.\ (\ref{eq:nonint}) and the model space in Eq.\ (\ref{eq:modelspace}) are not compatible since it is not possible to partition Eq.\ (\ref{eq:nonint}) in terms of the one-fermion model spaces. However, we can reorder the spinor elements of the wave function as
\begin{gather}
 \fett{P}^\mr{T}\fett{\Psi}(\fett{r})=\fett{\psi}_1(\fett{r}_1)\boxtimes\ldots\boxtimes\fett{\psi}_i(\fett{r}_i)\boxtimes\ldots\boxtimes\fett{\psi}_N(\fett{r}_N)
\end{gather}
where $\boxtimes$ is the Tracy--Singh product and $\fett{P}$ is a permutation matrix (see 
appendix \ref{sec:tsprod} for further details). Then, our two-spinor example reads
\begin{gather}
 \fett{\psi}_1(\fett{r}_1)\boxtimes\fett{\psi}_2(\fett{r}_2)=
\begin{bmatrix}
\highlight{\fett{\psi}^l_{1}(\fett{r}_1)\otimes\fett{\psi}^l_{2}(\fett{r}_2)}\\
\fett{\psi}^l_{1}(\fett{r}_1)\otimes\fett{\psi}^s_{2}(\fett{r}_2)\\
\fett{\psi}^s_{1}(\fett{r}_1)\otimes\fett{\psi}^l_{2}(\fett{r}_2)\\
\fett{\psi}^s_{1}(\fett{r}_1)\otimes\fett{\psi}^s_{2}(\fett{r}_2)\\
\end{bmatrix}=
\begin{bmatrix}
\highlight{\psi^{l1}_{1}(\fett{r}_1)\psi^{l1}_{2}(\fett{r}_2)}\\
\highlight{\psi^{l1}_{1}(\fett{r}_1)\psi^{l2}_{2}(\fett{r}_2)}\\
\highlight{\psi^{l2}_{1}(\fett{r}_1)\psi^{l1}_{2}(\fett{r}_2)}\\
\highlight{\psi^{l2}_{1}(\fett{r}_1)\psi^{l2}_{2}(\fett{r}_2)}\\
\psi^{l1}_{1}(\fett{r}_1)\psi^{s1}_{2}(\fett{r}_2)\\
\psi^{l1}_{1}(\fett{r}_1)\psi^{s2}_{2}(\fett{r}_2)\\
\vdots\\
\psi^{s2}_{1}(\fett{r}_1)\psi^{s2}_{2}(\fett{r}_2)
\end{bmatrix}\ .\label{eq:twots}
\end{gather}

The spinor components highlighted in Eq.\ (\ref{eq:twots}) are those contained within the $\mathcal{H}^{ll}$ model space as in Eq.\ (\ref{eq:twokr}). 
We see that the wave function in Eq.\ (\ref{eq:twots}) can be partitioned such that the individual components are part of the different model spaces in Eq.\ (\ref{eq:modelspace}),
\begin{gather}
 \left[\fett{P}^\mr{T}\fett{\Psi}(\fett{r})\right]^{\lambda_1\ldots \lambda_N}=\fett{\psi}^{\lambda_1}_1(\fett{r}_1)\otimes\ldots\otimes\fett{\psi}^{\lambda_i}_i(\fett{r}_i)\otimes\ldots\otimes\fett{\psi}^{\lambda_N}_N(\fett{r}_N)\ ,
\end{gather}
where $\lambda_1,\ldots,\lambda_N\in\left\{l,s\right\}$ as in Eq.\ (\ref{eq:lambda}) and antisymmetrization will be required.

The Hamiltonian is transformed accordingly (cf.\ Eq.\ (\ref{eq:mat}) in the appendix)
\begin{gather}
 \fett{H}^{(N)}_{DTS}=\fett{P}^\mr{T}\fett{H}^{(N)}_{D}\fett{P}
=\sum_{i=1}^N\fett{P}^\mr{T}\fett{h}_D(i)\fett{P}+\fett{P}^\mr{T}\fett{W}^{(N)}\fett{P}
\equiv \sum_{i=1}^N\fett{h}_{DTS}(i)+\fett{P}^\mr{T}\fett{W}^{(N)}\fett{P}\label{eq:nphamiltoniantrs}
\end{gather}
with
\begin{gather}
 \fett{h}_{DTS}(i)=\fett{P}^\mr{T}\fett{h}_D(i)\fett{P}=\fett{1}_4(1)\boxtimes\cdots\boxtimes\fett{1}_4(i-1)\boxtimes\fett{h}_D\boxtimes\fett{1}_4(i+1)\boxtimes\cdots\boxtimes\fett{1}_4(N).
\label{eq:freenphamiltoniantrs}
\end{gather}
The potential-energy operator $\fett{W}^{(N)}$ will be invariant under this transformation if only the instantaneous Coulomb interaction is considered as it is a diagonal matrix with identical entries. The situation is more complicated when magnetic interactions are taken into account. An $N$-fermion wave function for 1/2-fermions can then be partitioned in terms of the model space into $2^N$ components each of dimension $2^N$,
\begin{gather}
 \fett{\Psi}(\fett{r})=\begin{bmatrix}\fett{\Psi}^{l\ldots l}(\fett{r})\\\ldots\\\fett{\Psi}^{\lambda_1\ldots \lambda_N}(\fett{r})\\\ldots\\\fett{\Psi}^{s\ldots s}(\fett{r})\end{bmatrix}\label{eq:npwavefnctn}\ 
 =\begin{bmatrix}
| l \ldots l \rangle\\
\ldots\\
| \lambda_1\ldots \lambda_N \rangle\\
\ldots\\
| s \ldots s \rangle
\end{bmatrix}\
.
\end{gather}
Note that a related reordering of the Hamiltonian similar to Eq.\ (\ref{eq:nphamiltoniantrs}) is key for the quaternion formulation of four-component self-consistent field algorithms\cite{Saue1997}.

\section{Exact Two-Particle Kinetic-Balance Condition}
\label{sec:kb}

In this section, we derive the kinetic-balance condition for explicitly correlated basis functions for a system of two fermions. According to Eq.\ (\ref{eq:modelspace}) the model space takes the form
\begin{gather}
 \mathcal{H}^{(2)}=\mathcal{H}^{ll}\oplus\mathcal{H}^{ls}\oplus\mathcal{H}^{sl}\oplus\mathcal{H}^{ss},
\end{gather}
where the four subspaces
are formed from the single-fermion model spaces $\mathcal{H}^{l}$ and $\mathcal{H}^{s}$: 
\begin{gather}
 \mathcal{H}^{ll}=\mathcal{H}^{l}\otimes\mathcal{H}^{l}\label{eq:HLL},\\
 \mathcal{H}^{ls}=\mathcal{H}^{l}\otimes\mathcal{H}^{s}\label{eq:HLS},\\
 \mathcal{H}^{sl}=\mathcal{H}^{s}\otimes\mathcal{H}^{l}\label{eq:HSL},\\
 \mathcal{H}^{ss}=\mathcal{H}^{s}\otimes\mathcal{H}^{s}\label{eq:HSS}.
\end{gather}
Each model space in Eqs.\ (\ref{eq:HLL})--(\ref{eq:HSS}) is assigned to one of 
four components in the 16-component wave function. 
The structure of the Dirac Hamiltonian has to respect the structure of the Tracy--Singh product (see Eq.\ (\ref{eq:tracysingh}) in the appendix) to match 
the partitioning of the wave function according to Eq.\ (\ref{eq:npwavefnctn}). We then obtain the following block structure for the two-fermion Hamiltonian defined in Eq.\ (\ref{eq:nphamiltoniantrs}):
\begin{gather}
\hspace*{-1cm} \fett{H}^{(2)}_{DTS}(\fett{r}_1,\fett{r}_2)=\begin{bmatrix}\fett{V}+\fett{W}&c(\fett{\sigma}^{(2)}_2\cdot\fett{p}_2)&c(\fett{\sigma}^{(2)}_1\cdot\fett{p}_1)&\fett{0}_4\\
                           c(\fett{\sigma}^{(2)}_2\cdot\fett{p}_2)&\fett{V}+\fett{W}-2m_2c^2\fett{1}_4&\fett{0}_4&c(\fett{\sigma}^{(2)}_1\cdot\fett{p}_1)\\
              c(\fett{\sigma}^{(2)}_1\cdot\fett{p}_1)&\fett{0}_4&\fett{V}+\fett{W}-2m_1c^2\fett{1}_4&c(\fett{\sigma}^{(2)}_2\cdot\fett{p}_2)\\
              \fett{0}_4&c(\fett{\sigma}^{(2)}_1\cdot\fett{p}_1)&c(\fett{\sigma}^{(2)}_2\cdot\fett{p}_2)&\fett{V}+\fett{W}-2m_{12}c^2\fett{1}_4\\
\end{bmatrix}
\label{hmatrix}
\end{gather}
where we introduced the four-dimensional unit matrix $\fett{1}_4$ to highlight the dimension and 
$\fett{V}=[{V}(\fett{r}_1)+{V}(\fett{r}_2)]\fett{1}_4$ to yield a four-dimensional respresentation of the external potential-energy operator. 
Moreover, we assume that $\fett{V}$ and also the four-dimensional fermion--fermion interaction operator $\fett{W}$ are diagonal, 
which does not hold if magnetic and retardation effects are considered for the interaction of the two fermions (hence, we apply the compact
notation '$\fett{W}$' for a 4$\times$4 matrix operator describing the Coulomb interaction of two fermions only).
If this assumption is not made, rather complicated expressions will emerge for a magnetically balanced, explicitly correlated basis. In particular, the
zero entries in Eq.\ (\ref{hmatrix}) that represent the cases with a large and small component in the bracket per fermion would carry the magnetic fermion--fermion interaction 
(as expressed, for instance, in the Gaunt or Breit operators). As we will later make an assumption that all potential energy contributions are small compared to the rest energies
of the fermions, we aim at a kinetic balance condition free of any reference to a potential energy operator in analogy to the orbital-based two-fermion case.

Note that we have also introduced an energy shift of the whole spectrum in Eq.\ (\ref{hmatrix})
by $-m_{12}c^2$ with $m_{12}=m_1+m_2$. Moreover, we absorbed the direct products into $\fett{\sigma}^{(2)}_i$ as
\begin{gather}
\fett{\sigma}^{(2)}_1=\left(\fett{\sigma}_x\otimes\fett{1}_2,\fett{\sigma}_y\otimes\fett{1}_2,\fett{\sigma}_z\otimes\fett{1}_2\right)^\mr{T}\ ,
\end{gather}
and
\begin{gather}
\fett{\sigma}^{(2)}_2=\left(\fett{1}_2\otimes\fett{\sigma}_x,\fett{1}_2\otimes\fett{\sigma}_y,\fett{1}_2\otimes\fett{\sigma}_z\right)^\mr{T}\ .
\end{gather}

The idea of kinetic balance is to relate the small-component one-fermion model spaces to their large-component one-fermion model spaces in the eigenvalue problem
\begin{gather}
 \left(\fett{H}^{(2)}_{DTS}-E\fett{1}_{16}\right)\fett{\Psi}(\fett{r}_1,\fett{r}_2)=0\ .\label{eq:eigen}
\end{gather}
This leads to a system of four equations, analogously to Eqs.\ (\ref{eq:system1}) and (\ref{eq:system2}),
\begin{align}
 0&=(\fett{V}+\fett{W}-E\fett{1}_4)\fett{\Psi}^{ll}+c(\fett{\sigma}^{(2)}_2\cdot\fett{p}_2)\fett{\Psi}^{ls}+ c(\fett{\sigma}^{(2)}_1\cdot\fett{p}_1)\fett{\Psi}^{sl},\label{eq:4zero}\\
 0&=c(\fett{\sigma}^{(2)}_2\cdot\fett{p}_2)\fett{\Psi}^{ll}+(\fett{V}+\fett{W}-E\fett{1}_4-2m_2c^2\fett{1}_4)\fett{\Psi}^{ls}+ c(\fett{\sigma}^{(2)}_1\cdot\fett{p}_1)\fett{\Psi}^{ss},\label{eq:4one}\\
 0&=c(\fett{\sigma}^{(2)}_1\cdot\fett{p}_1)\fett{\Psi}^{ll}+(\fett{V}+\fett{W}-E\fett{1}_4-2m_1c^2\fett{1}_4)\fett{\Psi}^{sl}+ c(\fett{\sigma}^{(2)}_2\cdot\fett{p}_2)\fett{\Psi}^{ss},\label{eq:4two}\\
 0&=c(\fett{\sigma}^{(2)}_1\cdot\fett{p}_1)\fett{\Psi}^{ls}+c(\fett{\sigma}^{(2)}_2\cdot\fett{p}_2)\fett{\Psi}^{sl}+(\fett{V}+\fett{W}-E\fett{1}_4-2m_{12}c^2\fett{1}_4)\fett{\Psi}^{ss},\label{eq:4three}
\end{align}
where we have suppressed the coordinate dependence of the 4-spinors and will continue to do so where convenient.
We eliminate one of these four equations because we search for a relation between the four four-dimensional components of the wave function which we can then apply as a constraint 
on explicitly correlated basis functions. As in the case of a single fermion, we eliminate the energy $E$ from the equations by approximating 
\begin{gather}
\left[2m_ic^2+E\right]\fett{1}_4-\fett{V}-\fett{W}\approx2m_ic^2\fett{1}_4 \label{eq:approx}
\end{gather}
where $m_i\in\left\{m_1,m_2,m_1+m_2\right\}$. 
Similarly to the one-fermion case, Eq.\ (\ref{eq:oneferm}), we assume that this approximation remains valid and a variationally 
stable many-particle basis set can be derived. 

We eliminate the first equation, Eq.\ (\ref{eq:4zero}), from the system of equations since it is the only equation where $2m_ic^2$ does not occur so that Eq.\ (\ref{eq:approx}) cannot be applied. After applying Eq.\ (\ref{eq:approx}) to Eqs.\ (\ref{eq:4one})--(\ref{eq:4three}), we find the following relations among the four components of the two-fermion wave function:
\begin{align}
 0&\approx c(\fett{\sigma}^{(2)}_2\cdot\fett{p}_2)\fett{\Psi}^{ll}-2m_2c^2\fett{\Psi}^{ls}+ c(\fett{\sigma}^{(2)}_1\cdot\fett{p}_1)\fett{\Psi}^{ss}\label{eq:one},\\
 0&\approx c(\fett{\sigma}^{(2)}_1\cdot\fett{p}_1)\fett{\Psi}^{ll}-2m_1c^2\fett{\Psi}^{sl}+ c(\fett{\sigma}^{(2)}_2\cdot\fett{p}_2)\fett{\Psi}^{ss}\label{eq:two},\\
 0&\approx c(\fett{\sigma}^{(2)}_1\cdot\fett{p}_1)\fett{\Psi}^{ls}+ c(\fett{\sigma}^{(2)}_2\cdot\fett{p}_2)\fett{\Psi}^{sl}-2m_{12}c^2\fett{\Psi}^{ss}\label{eq:three}\ .
\end{align}
The matrix form of this under-determined system of linear equations can be interpreted as the augmented form of a linear system with a unique solution:
\begin{gather}
 \fett{A}=
 \begin{bmatrix}[ccc|c]
  (\fett{\sigma}^{(2)}_2\cdot\fett{p}_2)& -2m_2c\fett{1}_4&\fett{0}_4&-(\fett{\sigma}^{(2)}_1\cdot\fett{p}_1)\\
  (\fett{\sigma}^{(2)}_1\cdot\fett{p}_1)&\fett{0}_4& -2m_1c\fett{1}_4&-(\fett{\sigma}^{(2)}_2\cdot\fett{p}_2)\\
  \fett{0}_4&(\fett{\sigma}^{(2)}_1\cdot\fett{p}_1)&(\fett{\sigma}^{(2)}_2\cdot\fett{p}_2)&2m_{12}c\fett{1}_4\\
 \end{bmatrix}\begin{matrix}
  \{1\}\\\{2\}\\\{3\}
 \end{matrix}\label{eq:augmentedForm}
\end{gather}
The augmented form of linear systems and row reduction are explained in somewhat
more detail in appendix \ref{sec:rrm}. The number in curly brackets on the right-hand side counts every row. It will be used to express the manipulations in the row reduction below.

There is no row-reduced echelon form for the augmented form in Eq.\ (\ref{eq:augmentedForm}). The lack of a multiplicative inverse of the differential operator prohibits 
setting the leading element of each row of the row-reduced echelon form to 1 (see Eq.\ (\ref{eq:rreunity}) in the appendix) and therefore to relate $\Psi^{ll}(\fett{r}_1,\fett{r}_2)$, $\Psi^{ls}(\fett{r}_1,\fett{r}_2)$, and $\Psi^{sl}(\fett{r}_1,\fett{r}_2)$ to $\Psi^{ss}(\fett{r}_1,\fett{r}_2)$. However, we are able to find a similar form with pairwise relations between $\Psi^{ss}(\fett{r}_1,\fett{r}_2)$ and the other three components. 
These individual steps are to be taken in order to obtain this modified row-reduced echelon form:

\begin{enumerate}
 \item Insert $(\fett{\sigma}^{(2)}_1\cdot\fett{p}_1)\{1\}-(\fett{\sigma}^{(2)}_2\cdot\fett{p}_2)\{2\}$ into $\{2\}$:
\begin{fleqn}[0pt]
\begin{alignat}{2}
  &\begin{bmatrix}[ccc|c]
  (\fett{\sigma}^{(2)}_2\cdot\fett{p}_2)& -2m_2c\fett{1}_4&\fett{0}_4&-(\fett{\sigma}^{(2)}_1\cdot\fett{p}_1)\\
  \fett{0}_4&-2m_2c(\fett{\sigma}^{(2)}_1\cdot\fett{p}_1)&2m_1c(\fett{\sigma}^{(2)}_2\cdot\fett{p}_2)&\left[\fett{p}^2_2-\fett{p}^2_1\right]\fett{1}_4\\
  \fett{0}_4&(\fett{\sigma}^{(2)}_1\cdot\fett{p}_1)&(\fett{\sigma}^{(2)}_2\cdot\fett{p}_2)&2m_{12}c\fett{1}_4\\
 \end{bmatrix}\begin{matrix}
  \{1\}\\\{2\}\\\{3\}\nonumber
 \end{matrix}
\end{alignat}
\end{fleqn}
 \item Insert $\{2\}+2m_2c\{3\}$ into $\{3\}$:
\begin{fleqn}[0pt]
\begin{alignat}{2}
 \begin{bmatrix}[ccc|c]
  (\fett{\sigma}^{(2)}_2\cdot\fett{p}_2)& -2m_2c\fett{1}_4&\fett{0}_4&-(\fett{\sigma}^{(2)}_1\cdot\fett{p}_1)\\
  \fett{0}_4&-2m_2c(\fett{\sigma}^{(2)}_1\cdot\fett{p}_1)&2m_1c(\fett{\sigma}^{(2)}_2\cdot\fett{p}_2)&\left[\fett{p}^2_2-\fett{p}^2_1\right]\fett{1}_4\\
  \fett{0}_4&\fett{0}_4&2m_{12}c(\fett{\sigma}^{(2)}_2\cdot\fett{p}_2)&\fett{p}^2_2-\fett{p}^2_1+4m_2m_{12}c^2\fett{1}_4\\
 \end{bmatrix}\begin{matrix}
  \{1\}\\\{2\}\\\{3\}\nonumber
 \end{matrix}
\end{alignat}
\end{fleqn}
 \item Insert $\displaystyle-\frac{m_{12}}{m_2}\{2\}+\frac{m_1}{m_2}\{3\}$ into $\{2\}$:
\begin{fleqn}[0pt]
\begin{alignat}{2}
 \begin{bmatrix}[ccc|c]
  (\fett{\sigma}^{(2)}_2\cdot\fett{p}_2)& -2m_2c\fett{1}_4&\fett{0}_4&-(\fett{\sigma}^{(2)}_1\cdot\fett{p}_1)\\
  \fett{0}_4&2m_{12}c(\fett{\sigma}^{(2)}_1\cdot\fett{p}_1)&\fett{0}_4&\left[\fett{p}^2_1-\fett{p}^2_2+4m_1m_{12}c^2\right]\fett{1}_4\\
  \fett{0}_4&\fett{0}_4&2m_{12}c(\fett{\sigma}^{(2)}_2\cdot\fett{p}_2)&\left[\fett{p}^2_2-\fett{p}^2_1+4m_2m_{12}c^2\right]\fett{1}_4\\
 \end{bmatrix}\begin{matrix}
  \{1\}\\\{2\}\\\{3\}\nonumber
 \end{matrix}
\end{alignat}
\end{fleqn}
 \item Insert $\displaystyle(\fett{\sigma}^{(2)}_1\cdot\fett{p}_1)m_{12}\{1\}+m_2\{2\}$ into $\{1\}$:
\begin{fleqn}[0pt]
\begin{alignat}{2}
\hspace*{-2.5cm}\begin{bmatrix}[ccc|c]
  m_{12}(\fett{\sigma}^{(2)}_2\cdot\fett{p}_2)(\fett{\sigma}^{(2)}_1\cdot\fett{p}_1)& \fett{0}_4&\fett{0}_4&\left[-m_1\fett{p}^2_1-m_2\fett{p}^2_2+4m_1m_2m_{12}c^2\right]\fett{1}_4\\
  \fett{0}_4&2m_{12}c(\fett{\sigma}^{(2)}_1\cdot\fett{p}_1)&\fett{0}_4&\left[\fett{p}^2_1-\fett{p}^2_2+4m_1m_{12}c^2\right]\fett{1}_4\\
  \fett{0}_4&\fett{0}_4&2m_{12}c(\fett{\sigma}^{(2)}_2\cdot\fett{p}_2)&\left[\fett{p}^2_2-\fett{p}^2_1+4m_2m_{12}c^2\right]\fett{1}_4\\
 \end{bmatrix}\begin{matrix}
  \{1\}\\\{2\}\\\{3\}\nonumber
 \end{matrix}
\end{alignat}
\end{fleqn}
\end{enumerate}
We arrive at a set of simple pairwise relations between $\fett{\Psi}^{ss}(\fett{r}_1,\fett{r}_2)$ and the other three components
\begin{align}
 -(\fett{\sigma}^{(2)}_1\cdot\fett{p}_1)(\fett{\sigma}^{(2)}_2\cdot\fett{p}_2)m_{12}\fett{\Psi}^{ll}&=\left(m_1\fett{p}_1^2+m_2\fett{p}_2^2-4m_1m_2m_{12}c^2\right)\fett{\Psi}^{ss}\label{eq:uutoll},\\
-2c(\fett{\sigma}^{(2)}_1\cdot\fett{p}_1)m_{12}\fett{\Psi}^{ls}&=\left(\fett{p}_2^2-\fett{p}_1^2-4m_1m_{12}c^2\right)\fett{\Psi}^{ss}\label{eq:ultoll},\\
-2c(\fett{\sigma}^{(2)}_2\cdot\fett{p}_2)m_{12}\fett{\Psi}^{sl}&=\left(\fett{p}_1^2-\fett{p}_2^2-4m_2m_{12}c^2\right)\fett{\Psi}^{ss}\label{eq:lutoll}.
\end{align}
Forming the least common multiple from the operators
on the left-hand sides of the equations, we can introduce a four-dimensional spinor $\fett{\Theta}(\fett{r}_1,\fett{r}_2)$ related to the $\fett{\Psi}^{ss}(\fett{r}_1,\fett{r}_2)$ component,
\begin{gather}
\label{thetadef}
\fett{\Psi}^{ss}(\fett{r}_1,\fett{r}_2)=-2cm_{12}(\fett{\sigma}^{(2)}_1\cdot\fett{p}_1)(\fett{\sigma}^{(2)}_2\cdot\fett{p}_2)\fett{\Theta}(\fett{r}_1,\fett{r}_2)\ ,
\end{gather}
insert it into Eqs.\ (\ref{eq:uutoll})--(\ref{eq:lutoll}) and eliminate identical terms on both sides. Instead of relating the upper component to the lower component, 
we relate all four four-dimensional components of the 16-spinor,
\begin{gather}
\fett{\Psi}(\fett{r})=
\begin{bmatrix}\fett{\Psi}^{ll}(\fett{r})\\\fett{\Psi}^{ls}(\fett{r})\\\fett{\Psi}^{sl}(\fett{r})\\\fett{\Psi}^{ss}(\fett{r})\end{bmatrix}
=
\begin{bmatrix}
\ket{ll}\\\ket{ls}\\\ket{sl}\\\ket{ss}
\end{bmatrix}
\end{gather}
to a common spinor $\fett{\Theta}(\fett{r}_1,\fett{r}_2)$:
\begin{align}
\ket{ll}&
=\left(m_1\fett{p}_1^2+m_2\fett{p}_2^2-4m_1m_2m_{12}c^2\right){\bf 1}_4\fett{\Theta}(\fett{r}_1,\fett{r}_2)\label{eq:uu}
\equiv U^{(2)}_{ll}\fett{\Theta}(\fett{r}_1,\fett{r}_2)
, \\
\ket{ls}&
=\frac{(\fett{\sigma}^{(2)}_2\cdot\fett{p}_2)}{2c}\left(\fett{p}_2^2-\fett{p}_1^2-4m_1m_{12}c^2\right)\fett{\Theta}(\fett{r}_1,\fett{r}_2)\label{eq:ul}
\equiv U^{(2)}_{ls}\fett{\Theta}(\fett{r}_1,\fett{r}_2)
, \\
\ket{sl}&
=\frac{(\fett{\sigma}^{(2)}_1\cdot\fett{p}_1)}{2c}\left(\fett{p}_1^2-\fett{p}_2^2-4m_2m_{12}c^2\right)\fett{\Theta}(\fett{r}_1,\fett{r}_2)\label{eq:lu}
\equiv U^{(2)}_{sl}\fett{\Theta}(\fett{r}_1,\fett{r}_2)
, \\
\ket{ss}&
=-m_{12}(\fett{\sigma}^{(2)}_1\cdot\fett{p}_1)(\fett{\sigma}^{(2)}_2\cdot\fett{p}_2)\fett{\Theta}(\fett{r}_1,\fett{r}_2)\label{eq:ll}
\qquad\qquad\,
\equiv U^{(2)}_{ss}\fett{\Theta}(\fett{r}_1,\fett{r}_2).
\end{align}
Here, we have introduced the short-hand notation $U^{(2)}_{ll}$, $U^{(2)}_{ls}$, $U^{(2)}_{sl}$, and $U^{(2)}_{ss}$
for the transformation to kinetically balanced components in analogy to the one-fermion case in Eq.\ (\ref{eq:kbtrans}).
In a subsequent section, we refer to the $i$-th term in the prefactor of such expressions as $d_i^{(N)}$ with $N$=2 for the two-fermion case; e.g.,
$d_3^{(N)}$ for $\ket{sl}$ is then $-\fett{\sigma}^{(2)}_1\cdot\fett{p}_1/2c\times 4m_2m_{12}c^2$.
The physical role of  $\fett{\Theta}(\fett{r}_1,\fett{r}_2)$ will become clear when we study the non-relativistic limit (see below).
We emphasize that $\fett{\Theta}(\fett{r}_1,\fett{r}_2)$ is in general an explicitly correlated geminal rather than a simple orbital product.

Because of the derivation in Eq.\ (\ref{thetadef}), $\fett{\Psi}^{ss}(\fett{r}_1,\fett{r}_2)$ is uniquely defined by $\fett{\Theta}(\fett{r}_1,\fett{r}_2)$ up to a constant, i.e., the constant of integration. 
For square-integrable functions, this constant is zero. Hence, cancellation of differential operators is not a problem and all components are uniquely determined by $\fett{\Theta}(\fett{r}_1,\fett{r}_2)$.

Finally, we consider fermion exchange symmetry (Pauli principle) for the two identical fermions leading to the relations\cite{pest12}
\begin{gather}
\fett{\Psi}^{ll}(\fett{r}_1,\fett{r}_2)=-\fett{\Psi}^{ll}(\fett{r}_2,\fett{r}_1),\\
\fett{\Psi}^{ls}(\fett{r}_1,\fett{r}_2)=-\fett{\Psi}^{sl}(\fett{r}_2,\fett{r}_1),\\
\fett{\Psi}^{ss}(\fett{r}_1,\fett{r}_2)=-\fett{\Psi}^{ss}(\fett{r}_2,\fett{r}_1),
\end{gather}
which have to be fulfilled in addition to the relations in Eqs.\ (\ref{eq:uu})--(\ref{eq:ll}). $\fett{\Theta}(\fett{r}_1,\fett{r}_2)$ is antisymmetrized before the components are constructed according to Eqs.\ (\ref{eq:uu})--(\ref{eq:ll}) because the operators $(\fett{\sigma}^{(2)}_1\cdot\fett{p}_1)$ and $(\fett{\sigma}^{(2)}_2\cdot\fett{p}_2)$ do not commute with the permutation operator which exchanges fermions 1 and 2.

\section{The Non-Relativistic Limit}
\label{sec:nrlimit}
The one-fermion kinetic-balance condition yields the correct non-relativistic limit for $c\rightarrow \infty$. This is a key requirement ensuring variational stability. We therefore require any kinetic-balance condition for more than one fermion to yield the correct non-relativistic limit.

Finding the non-relativistic limit for the one-fermion case is fairly trivial. For the two-fermion kinetic-balance condition, this is somewhat more involved. In order to find the correct limit, we rely on de l'H\^{o}spital's rule for limits,
\begin{gather}
 \lim_{x\rightarrow y}\frac{f(x)}{g(x)}=\lim_{x\rightarrow y}\frac{f'(x)}{g'(x)},
\end{gather}
where $f'(x)$ and $g'(x)$ are the derivatives of $f(x)$ and $g(x)$ with respect to $x$, whereas $y$ is the limiting value of $x$.

The non-relativistic limit of the two-fermion total energy for a wave function kinetically balanced according to Eqs.\ (\ref{eq:uu})--(\ref{eq:ll}), can be taken as a limiting case of the Rayleigh quotient
\begin{gather}
\displaystyle E_{\mr{NR}}=\lim_{c\rightarrow \infty}\frac{\bra{\fett{\Psi}}\fett{H}^{(2)}_{DTS}\ket{\fett{\Psi}}}{\braket{\fett{\Psi}}{\fett{\Psi}}}.
\label{eq:rc}
\end{gather}
For the one-electron part in $\bra{\fett{\Psi}}\fett{H}^{(2)}_{DTS}\ket{\fett{\Psi}}$ we have
\begin{eqnarray}
\bra{\fett{\Psi}}\sum_{i=1}^2\fett{h}_{DTS}(\fett{r}_i)\ket{\fett{\Psi}}
&=&
 \bra{ll}c(\fett{\sigma}^{(2)}_2\cdot\fett{p}_2)\ket{ls}
+\bra{ll}c(\fett{\sigma}^{(2)}_1\cdot\fett{p}_1)\ket{sl}\nonumber\\
&&+\bra{ls}c(\fett{\sigma}^{(2)}_2\cdot\fett{p}_2)\ket{ll}
-\bra{ls}2m_2c^2\ket{ls}\nonumber\\
&&+\bra{ls}c(\fett{\sigma}^{(2)}_1\cdot\fett{p}_1)\ket{ss}
+\bra{sl}c(\fett{\sigma}^{(2)}_1\cdot\fett{p}_1)\ket{ll}\nonumber\\
&&-\bra{sl}2m_1c^2\ket{sl}
+\bra{sl}c(\fett{\sigma}^{(2)}_2\cdot\fett{p}_2)\ket{ss}\nonumber\\
&&+\bra{ss}c(\fett{\sigma}^{(2)}_1\cdot\fett{p}_1)\ket{ls}
+\bra{ss}c(\fett{\sigma}^{(2)}_2\cdot\fett{p}_2)\ket{sl}\nonumber\\
&&-\bra{ss}2m_{12}c^2\ket{ss}
+\bra{\fett{\Psi}}\fett{V}\otimes\fett{1}_4\ket{\fett{\Psi}},
\end{eqnarray}
where we have not resolved the potential-energy expectation value for convenience.
It must now be noted that
\begin{equation}
\bra{ls}c(\fett{\sigma}^{(2)}_2\cdot\fett{p}_2)\ket{ll} -\bra{ls}2m_2c^2\ket{ls} +\bra{ls}c(\fett{\sigma}^{(2)}_1\cdot\fett{p}_1)\ket{ss}=0,
\end{equation}
which can be shown by exploiting Eqs.\ (\ref{eq:uutoll}) and (\ref{eq:ultoll}) to replace $\ket{ll}$ and $\ket{ls}$ by expressions for $\ket{ss}$.
Analogously, we can exploit Eqs.\ (\ref{eq:uutoll})--(\ref{eq:lutoll}) to show 
\begin{eqnarray}
\bra{sl}c(\fett{\sigma}^{(2)}_1\cdot\fett{p}_1)\ket{ll}
-\bra{sl}2m_1c^2\ket{sl}
+\bra{sl}c(\fett{\sigma}^{(2)}_2\cdot\fett{p}_2)\ket{ss}=0,\\
\bra{ss}c(\fett{\sigma}^{(2)}_1\cdot\fett{p}_1)\ket{ls}
+\bra{ss}c(\fett{\sigma}^{(2)}_2\cdot\fett{p}_2)\ket{sl}
-\bra{ss}2m_{12}c^2\ket{ss}
=0.
\end{eqnarray}
Hence, we find for the full Hamiltonian with interacting fermions
\begin{gather}
\bra{\fett{\Psi}}\fett{H}^{(2)}_{DTS}\ket{\fett{\Psi}}
=
 \bra{ll}c(\fett{\sigma}^{(2)}_2\cdot\fett{p}_2)\ket{ls}
+\bra{ll}c(\fett{\sigma}^{(2)}_1\cdot\fett{p}_1)\ket{sl}
+\bra{\fett{\Psi}}(\fett{V}+\fett{W})\otimes\fett{1}_4\ket{\fett{\Psi}}.
\end{gather}
We now apply de l'H\^{o}spital's rule to Eq.\ (\ref{eq:rc}) by taking the fourth-order derivative with respect to $c$ of both the numerator and the denominator:
\begin{gather}
\displaystyle  E_{\mr{NR}}
= \lim_{c\rightarrow \infty}\displaystyle\frac{\displaystyle\frac{d^4}{d c^4}\left\{\bra{ll}c(\fett{\sigma}^{(2)}_1\cdot\fett{p}_1)\ket{ls}+\bra{ll}c(\fett{\sigma}^{(2)}_2\cdot\fett{p}_2)\ket{sl}+\bra{\fett{\Psi}}(\fett{V}+\fett{W})\otimes\fett{1}_4\ket{\fett{\Psi}}\right\}}{\displaystyle\frac{d^4}{d c^4}\braket{\fett{\Psi}}{\fett{\Psi}}}\nonumber\\[2ex]
=\lim_{c\rightarrow \infty}
\frac{\bra{\fett{\Theta}}192 m_{12}^2 m_1 m_2^2 \fett{p}_1^2 \fett{1}_4+ 192 m_{12}^2 m_1^2 m_2 \fett{p}_2^2\fett{1}_4+384 m_{12}^2 m_1^2 m_2^2(\fett{V}+\fett{W})+\mathcal{O}(c^{-2})\ket{\fett{\Theta}}}{384 m_{12}^2 m_1^2 m_2^2\braket{\fett{\Theta}}{\fett{\Theta}}}
\ .\label{eq:nrtmp}
\end{gather}

The potential energy term, $\fett{V}+\fett{W}$, may also contain contributions depending on $c$, but these contributions are of second or higher order in $c^{-1}$. 
When taking the limit, they are all zero and we find the limit to be a simplified Rayleigh quotient depending on $\fett{\Theta}(\fett{r}_1,\fett{r}_2)$
\begin{gather}
\label{2ferminter}
\displaystyle E_{\mr{NR}}=
\frac{\left\langle\fett{\Theta}\left|\displaystyle\frac{\fett{p}_1^2}{2m_1} +\displaystyle\frac{\fett{p}_2^2}{2m_2}+\fett{\tilde V}+\fett{\tilde W}\right|\fett{\Theta}\right\rangle}{\braket{\fett{\Theta}}{\fett{\Theta}}},
\end{gather}
where $\fett{\tilde V}$ and $\fett{\tilde W}$ are the limiting values with $c\rightarrow\infty$ for $\fett{V}$ and $\fett{W}$, respectively.
In Eq.\ (\ref{2ferminter}), we obtain the Schr\"{o}dinger energy and therefore the correct non-relativistic limit. 
The limit also identifies the four-dimensional spinor $\fett{\Theta}(\fett{r}_1,\fett{r}_2)$ as the non-relativistic two-fermion Schr\"{o}dinger wave function (note that this
function still features a four-dimensional spinor structure as it accounts for the spin of two electrons).

It is interesting to note that the value of the non-relativistic, $c\rightarrow\infty$, limit is determined by 
the leading terms in $c$ of the three components $\ket{ll}$, $\ket{ls}$, and $\ket{sl}$ in Eqs.\ (\ref{eq:uu})--(\ref{eq:lu}) define the non-relativistic limit when we apply de l'H\^{o}spital's rule. These leading terms are
\begin{align}
\label{tra1}
\ket{ll}(c^2)&:\quad -4m_1m_2m_{12}c^2\ \fett{\Theta},\\
\ket{ls}(c)&:\quad -2m_1m_{12}c(\fett{\sigma}^{(2)}_2\cdot\fett{p}_2) \fett{\Theta},
\end{align}
and
\begin{gather}
\label{tra2}
\ket{sl}(c):\quad -2m_2m_{12}c(\fett{\sigma}^{(2)}_1\cdot\fett{p}_1)\ \fett{\Theta}.
\end{gather}
We also note that Eqs.\ (\ref{tra1})--(\ref{tra2}) are related to Eq.\ (\ref{eq:oneekb}). If we apply Eq.\ (\ref{eq:oneekb}) for particles 1 and 2 subsequently to $\fett{\Theta}(\fett{r})$ and then multiply 
by $4m_1m_2m_{12}c^2$,
\begin{align}
\begin{bmatrix}
\ket{ll}(c^2)\\[1.2ex]
\ket{ls}(c)\\[1.2ex]
\ket{sl}(c)\\[1.2ex]
\ket{ss}(1)\\[1.2ex]
\end{bmatrix}
\rightarrow 4m_1m_2m_{12}c^2\begin{bmatrix}
\displaystyle \fett{\Theta}\\[1.2ex]
\displaystyle\frac{(\fett{\sigma}^{(2)}_2\cdot\fett{p}_2)}{2m_2c}\fett{\Theta}\\[1.2ex]
\displaystyle\frac{(\fett{\sigma}^{(2)}_1\cdot\fett{p}_1)}{2m_1c}\fett{\Theta}\\[1.2ex]
\displaystyle\frac{(\fett{\sigma}^{(2)}_1\cdot\fett{p}_1)(\fett{\sigma}^{(2)}_2\cdot\fett{p}_2)}{4m_1m_2c^2}\fett{\Theta}
\end{bmatrix}
\end{align}
we obtain the expressions of Eqs.\ (\ref{tra1})--(\ref{tra2}).
Hence, we have shown that the one-fermion kinetic-balance condition in Eq.\ (\ref{eq:oneekb}) is sufficient for obtaining the correct non-relativistic limit 
for a two-fermion system. 
At first sight, this seems reassuring as obtaining the correct non-relativistic limit has been connected to variational
stability for orbital-based theories (see, e.g., Ref.\ \cite{Dyall2012}).
However, the one-fermion kinetic-balance condition may not be sufficient to ensure variational stability in the general case
considered here\cite{Pestka2003,Pestka2004a,Pestka2007}. 
Accordingly, the non-relativistic limit will then not be a sufficient, albeit a necessary condition for variational stability.

\section{Kinetic-Balance Condition for More Than Two Fermions}
\label{sec:Ncase}
The derivation presented in section \ref{sec:kb} can also be applied to systems of more than two fermions, and thus establishes in its full form an exact kinetic-balance
condition for general (non-separable) $N$-particle basis functions. 
How such a generalization could be achieved for the approach of Pestka and co-workers\cite{Pestka2003,Pestka2004a,Pestka2007} is not obvious 
and was not discussed in their papers. In our ansatz, we obtain rather lengthy expressions for three fermions, which we refrain from presenting explicitly for the sake of brevity. The resulting expressions can, however, be expanded into a polynomial with respect to $c$. The individual terms $d^{(3)}_i(c)$ of the prefactor of the 3-fermion 8-spinor $\fett{\Theta}(\fett{r}_1,\fett{r}_2,\fett{r}_3)$ feature the important property
\begin{gather}
d^{(3)}_i(c)=k^{(3)}_i(m_1,m_2,m_3)\times c^{(6-u-v-w)}(\fett{\sigma}^{(3)}_1\cdot\fett{p}_1)^u(\fett{\sigma}^{(3)}_2\cdot\fett{p}_2)^v(\fett{\sigma}^{(3)}_3\cdot\fett{p}_3)^w\label{eq:terms}
\end{gather}
where we have omitted to indicate that each  $d^{(3)}_i(c)$ will be different for different sectors $|lll\rangle$, $|lls\rangle$, $|lss\rangle$, and so forth and
depend on $u,v,w$.
The positive semi-definite exponents $u,v,w$ obey the constraints $0\leq (u+v+w)\leq 7$ and we have
\begin{align}
\fett{\sigma}^{(3)}_1&=\left(\fett{\sigma}_x\otimes\fett{1}_4,\fett{\sigma}_y\otimes\fett{1}_4,\fett{\sigma}_z\otimes\fett{1}_4\right)^\mr{T}\ ,\label{eq:sp1}\\
\fett{\sigma}^{(3)}_2&=\left(\fett{1}_2\otimes\fett{\sigma}_x\otimes\fett{1}_2,\fett{1}_2\otimes\fett{\sigma}_y\otimes\fett{1}_2,\fett{1}_2\otimes\fett{\sigma}_z\otimes\fett{1}_2\right)^\mr{T}\ ,\label{eq:sp2}\\
\fett{\sigma}^{(3)}_3&=\left(\fett{1}_4\otimes\fett{\sigma}_x,\fett{1}_4\otimes\fett{\sigma}_y,\fett{1}_4\otimes\fett{\sigma}_z\right)^\mr{T}\ .\label{eq:sp3}
\end{align}
The multiplicative prefactors $k^{(3)}_i(m_1,m_2,m_3)$ depend on the masses of the individual fermions and the kinetic-balance conditions simplify significantly if all three fermions have equal masses.

Eq.\ (\ref{eq:terms}) shows that the explicitly correlated kinetic-balance condition for three particles contains 
the momentum operator to the power of seven, which is unfavorable from a computational point of view. However, we can observe
that the power of the momentum operators decreases with increasing orders of $c$. The leading terms with respect to $c$ are 
the one-fermion kinetic-balance terms and ensure the non-relativistic limit. 

For the assessment of the general properties of an $N$-fermion kinetic-balance condition, let us first re-write the two-fermion 
kinetic balance condition, Eqs.\ (\ref{eq:uu})--(\ref{eq:ll}), in a general form similar to Eq.\ (\ref{eq:terms}):
\begin{gather}
d^{(2)}_i(c)=k^{(2)}_i(m_1,m_2)\times c^{(4-u-v)}(\fett{\sigma}^{(2)}_1\cdot\fett{p}_1)^u(\fett{\sigma}^{(2)}_2\cdot\fett{p}_2)^v\label{eq:terms2}\ ,
\end{gather}
where the multiplicative prefactors $k^{(2)}_i(m_1,m_2)$ depend on the masses of the two fermions and the positive semi-definite exponents,
 $u$ and $v$, obey the constraints $0\leq (u+v) \leq 3$.

By comparing the results for two- and three-fermion systems, Eqs.\ (\ref{eq:terms}) and (\ref{eq:terms2}), 
we obtain for the $N$-fermion case: 
\begin{gather}
d^{(N)}_i(c)=k^{(N)}_i(m_1,\ldots,m_N)\times c^{(2N-u)}\prod_{j=1}^{N}(\fett{\sigma}^{(N)}_j\cdot\fett{p}_j)^{u_j}
\label{eq:termsN}
\end{gather}
where we skipped the explicit derivation. The power of $c$, $2N-u$, and the power of the $\fett{\sigma}^{(N)}_j\cdot\fett{p}_j$ operator, $u_j$,
are determined in the exact kinetic-balance solution by
\begin{gather}
\label{eq:termsNexp}
0\leq u=\sum_{j=1}^N u_j\leq 2N+1.
\end{gather}
High powers of the momentum operator is unfortunate from a computational point of view, but with the complete set of kinetic-balance conditions
at hand for any set of non-separable $N$-particle basis functions, Eqs.\ (\ref{eq:termsN}) and (\ref{eq:termsNexp}), 
one may introduce a hierarchy of approximate kinetic-balance conditions and
investigate their properties systematically.

As an example, we present the approximate kinetic-balance condition for a three-electron system (in Hartree atomic units and with
$m_\mathrm{e}=1$ for the electron mass) where only the leading terms in $c$ are included:
\begin{align}
\hspace*{-0.5cm}\ket{lll}&=\left(48 c^6\fett{1}_8 - 14 ((\fett{\sigma}^{(3)}_1\cdot\fett{p}_1)^2 + (\fett{\sigma}^{(3)}_2\cdot\fett{p}_2)^2 + (\fett{\sigma}^{(3)}_3\cdot\fett{p}_3)^2) c^4 \right)\fett{\Theta}(\fett{r})\label{eq:lll}
\equiv U^{(3)}_{lll}\fett{\Theta}(\fett{r}),\\
\hspace*{-0.5cm}\ket{lls}&=(\fett{\sigma}^{(3)}_3\cdot\fett{p}_3)\left((-(\fett{\sigma}^{(3)}_1\cdot\fett{p}_1)^2 - (\fett{\sigma}^{(3)}_2\cdot\fett{p}_2)^2 - 7 (\fett{\sigma}^{(3)}_3\cdot\fett{p}_3)^2) c^3 + 24 c^5\fett{1}_8\right)\fett{\Theta}(\fett{r})
\equiv U^{(3)}_{lls}\fett{\Theta}(\fett{r}),\\
\hspace*{-0.5cm}\ket{lsl}&=(\fett{\sigma}^{(3)}_2\cdot\fett{p}_2)\left((-(\fett{\sigma}^{(3)}_1\cdot\fett{p}_1)^2 - 7 (\fett{\sigma}^{(3)}_2\cdot\fett{p}_2)^2 - (\fett{\sigma}^{(3)}_3\cdot\fett{p}_3)^2) c^3 + 24 c^5\fett{1}_8\right)\fett{\Theta}(\fett{r})
\equiv U^{(3)}_{lsl}\fett{\Theta}(\fett{r}),\\
\hspace*{-0.5cm}\ket{lss}&=12 (\fett{\sigma}^{(3)}_2\cdot\fett{p}_2) (\fett{\sigma}^{(3)}_3\cdot\fett{p}_3) c^4\fett{\Theta}(\fett{r})
\equiv U^{(3)}_{lss}\fett{\Theta}(\fett{r}),\\
\hspace*{-0.5cm}\ket{sll}&=(\fett{\sigma}^{(3)}_1\cdot\fett{p}_1)\left((- 7 (\fett{\sigma}^{(3)}_1\cdot\fett{p}_1)^2 - (\fett{\sigma}^{(3)}_2\cdot\fett{p}_2)^2 - (\fett{\sigma}^{(3)}_3\cdot\fett{p}_3)^2) c^3 + 24 c^5\fett{1}_8\right)\fett{\Theta}(\fett{r})
\equiv U^{(3)}_{sll}\fett{\Theta}(\fett{r}),\\
\hspace*{-0.5cm}\ket{sls}&=12 (\fett{\sigma}^{(3)}_1\cdot\fett{p}_1) (\fett{\sigma}^{(3)}_3\cdot\fett{p}_3) c^4 \fett{\Theta}(\fett{r})
\equiv U^{(3)}_{sls}\fett{\Theta}(\fett{r}),\\
\hspace*{-0.5cm}\ket{ssl}&=12 (\fett{\sigma}^{(3)}_1\cdot\fett{p}_1) (\fett{\sigma}^{(3)}_2\cdot\fett{p}_2) c^4 \fett{\Theta}(\fett{r})
\equiv U^{(3)}_{ssl}\fett{\Theta}(\fett{r}),\\
\hspace*{-0.5cm}\ket{sss}&=-6 (\fett{\sigma}^{(3)}_1\cdot\fett{p}_1) (\fett{\sigma}^{(3)}_2\cdot\fett{p}_2) (\fett{\sigma}^{(3)}_3\cdot\fett{p}_3) c^3\fett{\Theta}(\fett{r})
\equiv U^{(3)}_{sss}\fett{\Theta}(\fett{r}) ,\label{eq:sss}
\end{align}
with the $\fett{\sigma}^{(3)}_i\cdot\fett{p}_i$ operators defined in Eqs.\ (\ref{eq:sp1})--(\ref{eq:sp2}). 
$\fett{\Theta}(\fett{r})$ with $\fett{r}=(\fett{r}_1,\fett{r}_2,\fett{r}_3)^{\mr{T}}$ is the non-relativistic limit of $\fett{\Psi}(\fett{r})$. We see that the lowest order of $c$ to consider is 3 due to the $\ket{sss}$ component. Eqs.\ (\ref{eq:lll})--(\ref{eq:sss}) can be 
considered as a minimal explicitly correlated kinetic-balance condition for a three-electron system.

\section{Basis-Set Expansion and Numerical Results}
\label{sec:num}
In practice, a many-particle wave function can be expanded into a basis set 
\begin{gather}
\fett{\Psi}(\fett{r})=\sum_{i}
\sum_{\lambda\in\Lambda}
C_i^{\lambda}
\fett{\Phi}_i^{\lambda}
(\fett{r})
\label{eq:trialwavefunction}
\end{gather}
where $C_i^{\lambda}$ are the expansion coefficients and $\fett{\Phi}_i^{\lambda}(\fett{r})$ are the basis functions.
$\Lambda$ is the set of all component-index strings consisting of $l$'s and $s$'s according to Eq.~(\ref{eq:npwavefnctn}), i.e., it is
the set of $2^N$ strings of such indices of length $N$ for an $N$-fermion basis function.
For the sake of clarity, we explicitly provide the basis functions for the two-fermion case,
\begin{gather}
\label{expanded}
\fett{\Phi}_i^{ll}=\begin{bmatrix} \fett{\tilde \Phi}_i^{ll}\\ 0\\ 0\\ 0 \end{bmatrix},~~
\fett{\Phi}_i^{ls}=\begin{bmatrix} 0\\ \fett{\tilde \Phi}_i^{ls}\\ 0\\ 0 \end{bmatrix},~~
\fett{\Phi}_i^{sl}=\begin{bmatrix} 0\\ 0\\ \fett{\tilde \Phi}_i^{sl}\\ 0 \end{bmatrix},~~\mbox{and}~~
\fett{\Phi}_i^{ss}=\begin{bmatrix} 0\\ 0\\ 0\\ \fett{\tilde \Phi}_i^{ss} \end{bmatrix},
\end{gather}
where the four-dimensional basis functions $\fett{\tilde \Phi}_i^{ll}$, $\fett{\tilde \Phi}_i^{ls}$, $\fett{\tilde \Phi}_i^{sl}$, and $\fett{\tilde \Phi}_i^{ss}$
are promoted to 16-dimensional functions for a compact notation of the expansion in Eq.\ (\ref{eq:trialwavefunction});
note that we write '0' in Eq.\ (\ref{expanded}) to indicate four-dimensional null vectors for the sake of brevity. Eventually, these four-dimensional basis functions
are to be expressed in terms of basis functions  $\fett{\Theta}_i$ that represent the common non-relativistic limit $\fett{\Theta}$ according to the analysis presented above.

A transformation, similar to that in Eq.\ (\ref{eq:kbtrans}) for the one-fermion case, can be formulated for the explicitly correlated kinetic-balance condition 
in the two-fermion case,
\begin{gather}
 \fett{U}^{(2)}_{\mr{KB}}=
\begin{bmatrix}
\displaystyle
\frac{
U^{(2)}_{ll}
}{|U^{(2)}_{ll}|}
&\fett{0}_4&\fett{0}_4&\fett{0}_4\\
\fett{0}_4&\displaystyle
\frac{
U^{(2)}_{ls}
}{|U^{(2)}_{ls}|}
&\fett{0}_4&\fett{0}_4\\
\fett{0}_4&\fett{0}_4&\displaystyle
\frac{
U^{(2)}_{sl}
}{|U^{(2)}_{sl}|}
&\fett{0}_4\\
\fett{0}_4&\fett{0}_4&\fett{0}_4&\displaystyle
\frac{
U^{(2)}_{ss}
}{|U^{(2)}_{ss}|}
\end{bmatrix},
\label{eq:kbtrans2}
\end{gather}
in the notation introduced in Eqs.\ (\ref{eq:uu})--(\ref{eq:ll}) and with a normalization introduced for each basis-function component according to
\begin{equation}
|U^{(2)}_{ll}|\equiv \sqrt{\langle \fett{\tilde\Phi}_i^{ll}\vert \fett{\tilde\Phi}_i^{ll} \rangle  } 
= \sqrt{\langle \fett{\Theta}_i\vert  U^{(2),\dagger}_{ll}\cdot U^{(2)}_{ll}   \vert\fett{\Theta}_i\rangle  } 
\end{equation}
and so forth for the other $\lambda$; note that we dropped the basis-function index on the left-hand side for the sake of brevity.
Essentially, we normalize each component of each basis function individually to ensure numerical stability when
solving the eigenvalue problem. This procedure can be understood as the relativistic counterpart of the quasi-normalization in pre-Born--Oppenheimer theory \cite{Matyus2012}.
Hence, explicit normalization of a trial wave function has to be taken into account when the energy is calculated.

In full analogy to the two-fermion case, we construct $\fett{U}^{(3)}_{\mr{KB}}$ from Eqs.\ (\ref{eq:lll})--(\ref{eq:sss}).
In general, the $N$-fermion trial wave function is expressed in terms of the transformation as
\begin{gather}
\fett{\Psi}(\fett{r})=\sum_{i} 
\begin{bmatrix}
C_i^{ll\dots l} \frac{\displaystyle {U}^{(N)}_{ll\dots l}}{\displaystyle |{U}^{(N)}_{ll\dots l}|}\fett{\Theta}_i(\fett{r})\\
\vdots\\
C_i^{\lambda} ~~~\frac{\displaystyle {U}^{(N)}_{\lambda}}{\displaystyle |{U}^{(N)}_{\lambda}|}\fett{\Theta}_i(\fett{r})\\
\vdots\\
C_i^{ss\dots s} \frac{\displaystyle {U}^{(N)}_{ss\dots s}}{\displaystyle |{U}^{(N)}_{ss\dots s}|}\fett{\Theta}_i(\fett{r})
\end{bmatrix}
\label{eq:trialwavefunctionN}
\equiv
\sum_{i} \fett{C}_i \cdot [\fett{U}^{(N)}_{\mr{KB}} \cdot\fett{\Theta}^{(N)}_i(\fett{r})]
\ ,
\end{gather}
where the ${U}^{(N)}_{\lambda}/|{U}^{(N)}_{\lambda}|$ are the entries of the diagonal matrix $\fett{U}^{(N)}_{\mr{KB}}$
normalized by 
\begin{equation}
|{U}^{(N)}_{\lambda}| \equiv \sqrt{\langle \fett{\tilde\Phi}^{\lambda}_i\vert \fett{\tilde\Phi}^{\lambda}_i\rangle}
=\sqrt{\langle \fett{\Theta}_i\vert  U^{(N),\dagger}_{\lambda} \cdot U^{(N)}_{\lambda}   \vert\fett{\Theta}_i\rangle}
\end{equation}
(with the index $i$ dropped for the sake of brevity as before).
The vector $\fett{\Theta}^{(N)}_i$ contains the non-relativistic limit, $\fett{\Theta}_i$, $2^N$ times as 
entry, i.e., $\fett{\Theta}^{(N)}_i=(\fett{\Theta}_i,\fett{\Theta}_i,\dots,\fett{\Theta}_i)$.

\subsection{Numerical Results}

As an example, we present numerical results for a standard two-electron system: 
two electrons moving in the central potential of a helium nucleus within the Born--Oppenheimer approximation.
Our starting point is a non-relativistic basis set, which corresponds to $L=0$ total spatial angular momentum, $p=+1$ parity, and
$S=0$ total electron spin quantum numbers, and which is antisymmetrizd according to the Pauli principle:
\begin{equation}
\label{pauli}
\fett{\Theta}_i(\fett{r})=\fett{\Theta}'_i(\fett{r}_1,\fett{r}_2) 
\frac{1}{\sqrt{2}} \left[
\left(\begin{array}{c}
1 \\ 0 \end{array}\right)
\otimes
\left(\begin{array}{c}
0 \\ 1 \end{array}\right)
-
\left(\begin{array}{c}
0 \\ 1 \end{array}\right)
\otimes
\left(\begin{array}{c}
1 \\ 0 \end{array}\right)
\right] ,
\end{equation}
where we inroduced the explicit form of the spin functions
\begin{equation}
\alpha=\left(\begin{array}{c}
1 \\ 0 \end{array}\right)
\quad\mbox{and}\quad
\beta=\left(\begin{array}{c}
0 \\ 1 \end{array}\right)  .
\end{equation}
In this notation, the four-dimensional structure of the non-relativistic limit is highlighted in agreement with Eq.\ (\ref{2ferminter}).
In Eq.\ (\ref{pauli}), the spatial part can be any non-separable two-particle function and in our calculations it is an explicitly
correlated Gaussian function with $L=0$ and $p=+1$,
\begin{gather}
\label{basisfct}
\fett{\Theta}'_i(\fett{r}_1,\fett{r}_2)=\exp\left(-\frac{1}{2}\fett{r}^{\mr{T}}\left(\fett{A}_i\otimes\fett{1}_3\right)\fett{r}\right)\ ,
\end{gather}
where $\fett{r}=(\fett{r}_1,\fett{r}_2)^{\mr{T}}$ and the elements of the symmetric, positive definite matrix, $\fett{A}_{i}\in\mathbb{R}^{2\times 2}$,
are parametrized by
\begin{gather}
\left\{\fett{A}_{i}\right\}_{kl}=\delta_{kl}\exp\left({\alpha}_{kl,i}\right)+0.1(\delta_{kl}-1)\exp\left(-{\alpha}_{kl,i}\right)\quad \mbox{with}~k,l\in\{1,2\}.
\end{gather}
The ${\alpha}_{kl,i}$ values are optimized stochastically to minimize the relativistic energy. Trial values for ${\alpha}_{kl,i}$ were generated
from a normal distribution as in Ref.\ \cite{Matyus2012} (see also references therein).
The optimized parameter values of $\fett{A}_i$ are deposited in the supplementary information.

With Eq.\ (\ref{eq:trialwavefunctionN}) (see also Eqs.\ (\ref{eq:uu})--(\ref{eq:ll}) for the two-particle case) we generate a kinetically balanced 
basis set from $\fett{\Theta}_i(\fett{r})$, Eq.\ (\ref{pauli}), for the relativistic calculations and
minimize the  Rayleigh quotient, Eq.\ (\ref{eq:E}),
\begin{gather}
 E[\fett{H}_{DTS}^{(2)},\left\{\fett{\Theta}_i(\fett{r})\right\}]=\frac{\sum_{ij} \fett{C}_i^\star \fett{C}_j\bra{\fett{U}^{(2)}_{\mr{KB}}\cdot\fett{\Theta}^{(2)}_i(\fett{r})}\fett{H}_{DTS}^{(2)}\ket{\fett{U}^{(2)}_{\mr{KB}}\cdot\fett{\Theta}^{(2)}_j(\fett{r})}}{\sum_{ij} \fett{C}_i^\star \fett{C}_j\braket{\fett{U}^{(2)}_{\mr{KB}}\cdot\fett{\Theta}^{(2)}_i(\fett{r})}{\fett{U}^{(2)}_{\mr{KB}}\cdot\fett{\Theta}^{(2)}_j(\fett{r})}}\label{eq:ray2}
\end{gather}
with respect to the expansion coefficients $C_i^{\lambda}$ by solving the generalized eigenvalue problem
\begin{gather}
\label{eigprob}
\fett{H}\fett{C}=\fett{E}\fett{S}\fett{C}\ .
\end{gather} 
In Eq.\ (\ref{eigprob}), the
Hamiltonian matrix, $\fett{H}$, has a block structure with
$\fett{H}_{ij}=\bra{\fett{U}^{(2)}_{\mr{KB}}\cdot\fett{\Theta}^{(2)}_i(\fett{r})}\fett{H}_{DTS}^{(2)}\ket{\fett{U}^{(2)}_{\mr{KB}}\cdot\fett{\Theta}^{(2)}_j(\fett{r})} \in \mathbb{R}^{2^N\times 2^N} ~(i,j=1,2,\dots,n)$ 
and similarly the overlap matrix, $\fett{S}$, contains
$\fett{S}_{ij}=\braket{\fett{U}^{(2)}_{\mr{KB}}\cdot\fett{\Theta}^{(2)}_i(\fett{r})}{\fett{U}^{(2)}_{\mr{KB}}\cdot\fett{\Theta}^{(2)}_j(\fett{r})}  \in \mathbb{R}^{2^N\times 2^N} ~(i,j=1,2,\dots,n)$ 
for $n$ basis functions and for two electrons, $N$=2. Accordingly,
$\fett{C} \in \mathbb{R}^{n2^N\times n2^N}$ is a matrix containing the expansions coefficients $C_i^{\lambda}$ and $\fett{E}$ is an $(n2^N)$-dimensional
diagonal matrix with the energies on its diagonal.

\begin{table}[h!]
 \caption{\label{tab:energies}\small Ground-state energy of the two-electron helium atom with fixed nucleus obtained for increasing basis-set sizes.
$\Delta E_{\mr{R}}$ and $\Delta E_{\mr{NR}}$ are the differences 
between the calculated relativistic and non-relativistic energies, respectively, with respect to the reference values. $n$ is the number of basis functions,
 $\fett{\Theta}_i$, defined in Eqs.\ (\ref{pauli})
and (\ref{basisfct}). The parameters of the basis functions are deposited in the Supplementary Material and the value for the speed of light was set to  137.0359895 atomic units.}
 \begin{center}
 \renewcommand{\baselinestretch}{1.0}
 \renewcommand{\arraystretch}{1}
\begin{tabular}{lllll}
\hline
\hline
$n$&$E_{\mr{R}}$ [$E_{h}$]&$\Delta E_{\mr{R}}$ [$E_{h}$]&$E_{\mr{NR}}$ [$E_{h}$]&$\Delta E_{\mr{NR}}$ [$E_{h}$]\\
\hline
10& -2.89757665&0.00628019&-2.89744422&0.00628016\\
20& -2.90288205&0.00097479&-2.90275061&0.00097377\\
50& -2.90382266&0.00003418&-2.90369103&0.00003335\\
100&-2.90384822&0.00000862&-2.90372140&0.00000298\\
200&-2.90385566&0.00000118&-2.90372429&0.00000009\\
300&-2.90385674&0.00000010&-2.90372430&0.00000008\\
\hline
Ref.\ \cite{Pestka2007}&-2.90385684&Ref.\ \cite{Freund1984}&-2.90372438\\
\hline
\hline
  \end{tabular}
 \end{center}
\end{table}

The ground-state energy eigenvalue of the helium atom is obtained from Eq.\ (\ref{eigprob})
by direct solution of the generalized eigenvalue problem in the stochastically optimized basis set
(see Table \ref{tab:energies}). The non-relativistic energies, also given in Table \ref{tab:energies}, were obtained from
the generalized eigenvalue problem solved for the Schr\"odinger Hamiltonian in the basis of the non-relativistic basis functions
of Eq.\ (\ref{pauli}), containing the parameters obtained in the relativistic calculations (see the supporting information for
details).
As it can be seen from the data in Table \ref{tab:energies}, both the relativistic and the non-relativistic energies converge with increasing basis-set size
towards the reference data in a variationally stable fashion.

\section{Conclusions}
The kinetic-balance condition for the one-fermion case ensures variational stability in orbital-based approaches to first-quantized relativistic many-fermion theory. 
In the present work, we derived a kinetic-balance condition for 
general, non-separable $N$-particle basis functions. Similarly to the derivation of a 
one-particle kinetic-balance condition, we set out from the assumption that 
the potential energy contributions are small compared to the rest 
energies of the fermions. We arrived at an $N$-particle kinetic balance 
condition by combining the well-known multiplication properties of the 
Pauli matrices with the row-elimination approach of solving linear 
systems of equations. In agreement with the one-fermion case, the 
$N$-particle kinetic-balance condition also ensures that the 
correct non-relativistic limit is obtained for an infinite speed of 
light. It had been anticipated, 
however, that the $N$-particle kinetic-balance condition 
provides better stability when solving the first-quantized Dirac Hamiltonian 
variationally with an explicitly correlated basis set, and hence 
suggested that the requirement of matching the non-relativistic limit 
is a necessary but not a sufficient condition.

We demonstrated that the variational solution of the Dirac equation is 
stable for the ground state of the two-fermion helium atom when a 
relativistic basis set is generated from explictily correlated 
Gaussian functions using the $N$-particle kinetic balance condition for $N$=2.

Concerning the general applicability of our results, the theoretical 
expressions and our preliminary investigations show that the direct 
use of the full $N$-particle kinetic-balance condition becomes tedious 
and computationally expensive for more than two fermions. However, it 
might be possible to reduce the computational cost by systemtically eliminating 
terms of high order in momentum operators from the exact expressions 
and, at the same time, retain variational stability for the 
solutions. A systematic investigation of the variational 
stability under such approximations is beyond the scope of the present paper 
and left for future work.

\section{Acknowledgments}
This work has been supported by the Swiss National Science Foundation SNF (project 200020\_156598). EM thanks the Hungarian Scientific Research Fund (OTKA, NK83583) for financial support.

\appendix
\section{Appendix}

\subsection{Tracy--Singh Product}
\label{sec:tsprod}
The Tracy--Singh product\cite{Tracy1972} is defined as
\begin{gather}
 \fett{A}_{\mr{tsp}}=\fett{B}\boxtimes\fett{C}=\left[\left(\fett{B}_{ij}\otimes\fett{C}_{uv}\right)\right]=
\begin{bmatrix}
\left(\fett{B}_{11}\otimes\fett{C}_{uv}\right)&\cdots&\left(\fett{B}_{1n}\otimes\fett{C}_{uv}\right)\\
\vdots&&\\
\left(\fett{B}_{m1}\otimes\fett{C}_{uv}\right)&\cdots&\left(\fett{B}_{mn}\otimes\fett{C}_{uv}\right)\\
\end{bmatrix}
\label{eq:tracysingh}
\end{gather}
where $\fett{B}=[\fett{B}_{ij}]$ and $\fett{C}=[\fett{C}_{uv}]$ are two matrices of dimension $(m\times n)$ and $(p\times q)$, respectively. They are partitioned block-wise in terms of the matrices $\fett{B}_{ij}$ and $\fett{C}_{uv}$. $\fett{A}_{\mr{tsp}}$ is a matrix of dimension $(mp\times nq)$. It is partitioned block-wise with the elements being the matrices $\left(\fett{B}_{ij}\otimes\fett{C}_{uv}\right)$. The Tracy--Singh product may be considered a more general form of the Kronecker product
\begin{gather}
\fett{A}_{\mr{kp}}=\fett{B}\otimes\fett{C}=\left[\left(b_{ij}c_{uv}\right)\right]=
\begin{bmatrix}
b_{11}\fett{C}&\cdots&b_{1n}\fett{C}\\
\vdots&&\\
b_{m1}\fett{C}&\cdots&b_{mn}\fett{C}\\
\end{bmatrix}
\end{gather}
where $b_{ij}$ and $c_{uv}$ are the matrix elements of $\fett{B}$ and $\fett{C}$, respectively. $\fett{A}_{\mr{kp}}$ is a matrix of dimension $(mp\times nq)$. The two matrices $\fett{A}_{\mr{tsp}}$ and $\fett{A}_{\mr{kp}}$ are identical in the case that $\fett{B}$ and $\fett{C}$ are not partitioned (or partitioned into $(1\times 1)$ blocks). Generally the two products are related through a permutation of the row and column space of either matrix\cite{Wei2000,Horn1992,Wansbeek1991}
\begin{gather}
 \fett{P}^T(\fett{B}_1\otimes\ldots\otimes\fett{B}_n)\fett{Q}=\fett{B}\boxtimes\ldots\boxtimes\fett{B}_n\label{eq:mat}
\end{gather}
where $\fett{P}$ and $\fett{Q}$ are the permutation matrices for the row and the column space and $n$ is the number of matrices involved. For vectors $\fett{v}_i$, we find the relation
\begin{gather}
 \fett{P}^\mr{T}(\fett{v}_1\otimes\ldots\otimes\fett{v}_n)=\fett{v}_1\boxtimes\ldots\boxtimes\fett{v}_n.
\label{eq:vect}
\end{gather}
The partitioning of the matrices and vectors depends on the permutation matrices $\fett{P}$ and $\fett{Q}$. If all matrices are square and symmetrically partitioned, the two permutation matrices are identical\cite{Wei2000}
 $\fett{P}=\fett{Q}$ and the two products are related through a unitary transformation.

\subsection{Row Reduction and Row Reduced Echelon Form}
\label{sec:rrm}
Systems of linear equations are conveniently solved by first representing them in matrix form
\begin{gather}
 \fett{A}\cdot\fett{x}-\fett{b}=0\ ,
\end{gather}
where $\fett{A}$ is a matrix containing the linear factors. $\fett{x}$ is a vector and contains the values which are to be determined and $\fett{b}$ is a vector containing the constant factors of the linear system. A reliable method of solving such a linear system is row reduction, i.e., Gaussian elimination. It involves performing a series of operations on the augmented form
\begin{gather}
 \fett{A}_{\mathrm{aug}}=\begin{bmatrix}
  \fett{A}|\fett{b}\\
 \end{bmatrix}
\end{gather}
until it is in row-reduced echelon form. The row-reduced echelon form is
\begin{gather}
 \fett{A}_{\mathrm{rre}}=\begin{bmatrix}
  \fett{1}|\fett{b}'\\
 \end{bmatrix}\label{eq:rreunity}
\end{gather}
for systems with a unique solution. Possible operations are permutation of two rows, multiplication of individual rows with a constant scalar factor and evaluating the difference of two rows.

\small

\end{document}